\documentclass[
prl,
aps,
 amsmath,amssymb,
reprint,
superscriptaddress,
floatfix,
]{revtex4-1}

\usepackage{graphicx}
\usepackage{dcolumn}
\usepackage{bm}

\usepackage{booktabs}
\usepackage{wasysym}
\usepackage[uncertainty-mode = separate]{siunitx}
\usepackage[utf8]{inputenc}
\usepackage[T1]{fontenc}
\usepackage{mathptmx}
\usepackage{etoolbox}
\usepackage{hyperref}

\hypersetup{
    colorlinks = true,
    linkcolor = blue,
    citecolor = blue,
    urlcolor=blue,
}

\usepackage{tabularx}
\usepackage{appendix}
\usepackage{caption}
\makeatletter
\def\@email#1#2{%
 \endgroup
 \patchcmd{\titleblock@produce}
  {\frontmatter@RRAPformat}
  {\frontmatter@RRAPformat{\produce@RRAP{*#1\href{mailto:#2}{#2}}}\frontmatter@RRAPformat}
  {}{}
}%
\makeatother

\usepackage[font=small,labelfont=bf,format=plain]{caption}

\usepackage{setspace}
  \usepackage{parskip}
\usepackage[compact]{titlesec}         
\titlespacing{\section}{4pt}{4pt}{4pt} 
\AtBeginDocument{
  \setlength\abovedisplayskip{4pt}
  \setlength\belowdisplayskip{4pt}
  }
  
\usepackage[]{caption, subfig}
\usepackage{lineno}
\setcitestyle{square}
\begin{document}
\newcommand{\ra}[1]{\renewcommand{\arraystretch}{#1}}

\title{Plasmoid formation and strong radiative cooling in a driven magnetic reconnection experiment}
\author{R. Datta}
\affiliation{Plasma Science and Fusion Center, Massachusetts Institute of Technology, MA 02139, Cambridge, USA\looseness=-10000 
}%
\author{K. Chandler}
\affiliation{ 
Sandia National Laboratories, Albuquerque, NM 87123-1106, USA
}%
\author{C.E. Myers}
\altaffiliation[Current address: ]{Commonwealth Fusion Systems, Devens, MA 01434, USA}
\affiliation{ 
Sandia National Laboratories, Albuquerque, NM 87123-1106, USA
}

\author{J. P. Chittenden}
\affiliation{ 
Blackett Laboratory, Imperial College London, London SW7 2BW, UK\looseness=-10000 
}%
\author{A. J. Crilly}
\affiliation{ 
Blackett Laboratory, Imperial College London, London SW7 2BW, UK\looseness=-10000 
}%

\author{C. Aragon}
\affiliation{ 
Sandia National Laboratories, Albuquerque, NM 87123-1106, USA
}%

\author{D. J. Ampleford}
\affiliation{ 
Sandia National Laboratories, Albuquerque, NM 87123-1106, USA
}%

\author{J. T. Banasek}
\affiliation{ 
Sandia National Laboratories, Albuquerque, NM 87123-1106, USA
}%

\author{A. Edens}
\affiliation{ 
Sandia National Laboratories, Albuquerque, NM 87123-1106, USA
}%
\author{W. R. Fox}
\affiliation{ 
Princeton Plasma Physics Laboratory, Princeton, NJ 08543, USA 
}%
\author{S. B. Hansen}
\affiliation{ 
Sandia National Laboratories, Albuquerque, NM 87123-1106, USA
}%
\author{E. C. Harding}
\affiliation{ 
Sandia National Laboratories, Albuquerque, NM 87123-1106, USA
}%
\author{C. A. Jennings}
\affiliation{ 
Sandia National Laboratories, Albuquerque, NM 87123-1106, USA
}%
\author{H. Ji}
\affiliation{ 
Princeton Plasma Physics Laboratory, Princeton, NJ 08543, USA 
}%
\author{C. C. Kuranz}
\affiliation{Department of Nuclear Engineering and Radiological Sciences, University of Michigan, Ann Arbor, MI 48109, USA\looseness=-1 
}%
\author{S. V. Lebedev}
\affiliation{ 
Blackett Laboratory, Imperial College London, London SW7 2BW, UK\looseness=-10000 
}%

\author{Q. Looker}
\affiliation{ 
Sandia National Laboratories, Albuquerque, NM 87123-1106, USA
}%

\author{S. G. Patel}
\affiliation{ 
Sandia National Laboratories, Albuquerque, NM 87123-1106, USA
}%

\author{A. Porwitzky}
\affiliation{ 
Sandia National Laboratories, Albuquerque, NM 87123-1106, USA
}%

\author{G. A. Shipley}
\affiliation{ 
Sandia National Laboratories, Albuquerque, NM 87123-1106, USA
}%

\author{D. A. Uzdensky}
\affiliation{Center for Integrated Plasma Studies, Physics Department, UCB-390, University of Colorado, Boulder, Colorado, USA\looseness=-1 
}%

\author{D. A. Yager-Elorriaga}
\affiliation{ 
Sandia National Laboratories, Albuquerque, NM 87123-1106, USA
}%

\author{J.D. Hare}%
\thanks{jdhare@mit.edu}
\affiliation{Plasma Science and Fusion Center, Massachusetts Institute of Technology, MA 02139, Cambridge, USA\looseness=-10000 
}%



\begin{abstract}
We present results from the first experimental study of strongly radiatively-cooled magnetic reconnection. Two exploding aluminum wire arrays, driven simultaneously by the Z machine ($I_{max} = 20 \, \text{MA}$, $t_{\text{rise}} = \SI{300}{\nano \second}$), generate a radiatively-cooled reconnection layer ($S_L \approx 120$) in which the total cooling rate exceeds the hydrodynamic transit rate ($\tau_{\text{hydro}}/\tau_{\text{cool}} > 100$). Measurements of X-ray emission from the reconnection layer using a filtered diode ($>1$ keV) show a narrow (50~ns FWHM) burst of emission at $\SI{220}{\nano \second}$ after current start, consistent with the formation and subsequent rapid cooling of the reconnection layer. Time-gated X-ray images of the reconnection layer show fast-moving (up to 50~km/s) hotspots inside the layer, consistent with the presence of plasmoids observed in 3D resistive magnetohydrodynamic simulations. X-ray spectroscopy shows that these hotspots generate the majority of Al K-shell emission (at around 1.6 keV) prior to the onset of cooling, and exhibit temperatures of \SI{170}{\electronvolt}, much greater than the temperature of the plasma inflows and the rest of the reconnection layer.

\end{abstract}

\maketitle

Magnetic reconnection is a ubiquitous process in magnetized plasmas, responsible for the explosive conversion of magnetic energy into heat and kinetic energy \cite{parker1957sweet,yamada2010magnetic,ji2022magnetic}. In extreme high-energy-density (HED) astrophysical systems, such as black hole accretion disks and their coronae \cite{goodman2008reconnection, beloborodov2017radiative,werner2019particle,mehlhaff2021pair}, pulsar magnetospheres \cite{lyubarsky2001reconnection,uzdensky2014physical,cerutti2015particle,schoeffler2019bright},  and gamma-ray bursts \cite{spruit2001large, uzdensky2007magnetically, giannios2008prompt,uzdensky2011magneticb,mckinney2012reconnection}, strong radiative cooling (e.g.\ inverse external Compton or synchrotron cooling) modifies the energy balance, by removing internal energy faster than it is injected into the reconnection layer \cite{uzdensky2011magnetic,uzdensky2011magneticb,uzdensky2016radiative}. This can trigger the radiative collapse of the layer ---  a runaway cooling and compression process --- which generates a cold, thin, and dense current sheet, that is theoretically predicted to increase the reconnection rate \cite{dorman1995one,uzdensky2011magnetic}.

Despite the importance of radiative cooling in astrophysical systems, there has been limited investigation of radiatively-cooled reconnection in laboratory plasmas, as it is difficult to achieve the cooling rates necessary to observe significant cooling on experimental time scales. Previous experiments on magnetically-driven devices, such as MRX $(n_e \sim \SI{e13}{\per \centi \meter \cubed}, \, T_e \sim 10 \, \text{eV}, \, \beta \ll 1)$ \cite{yamada1997study, yamada2010magnetic}, have provided valuable insight into reconnection physics in a low-density regime where radiative cooling is negligible, including evidence for Sweet-Parker reconnection \cite{yamada1997study,ji1999magnetic}, strong ion heating \cite{hsu2000local}, and the Hall effect \cite{ren2005experimental}. In contrast, laser-driven experiments operate in a strongly-driven $\beta \gg 1$ HED regime $(n_e \sim \SI{e20}{\per \centi \meter \cubed}, \, T_e \sim 1000 \,\, \text{eV})$ \cite{rosenberg2015slowing}, and have provided evidence for two-fluid effects, magnetic flux pile-up, and particle acceleration  \cite{rosenberg2015slowing, nilson2006magnetic,li2007observation,fiksel2014magnetic,chien2023non}. Despite the high densities and temperatures in these laser-driven experiments, the cooling parameter $R_{\text{cool}} \equiv \tau_{\text{hydro}}/\tau_{\text{cool}}$ was small, as the plasma ions in the reconnection layer have no bound electrons at these high temperatures \cite{fox2012magnetic,fiksel2014magnetic}, limiting cooling by line emission. A third type of reconnection experiment, driven by pulsed-power-driven, provide access to reconnection in a strongly-driven $\beta \approx 1$ HED regime \cite{lebedev2019exploring}. Experiments on the MAGPIE facility $(n_e \sim \SI{e18}{\per \centi \meter \cubed}, \, T_e \sim 50 \,\text{eV})$ drove a $\SI{1.4}{\mega\ampere}$ current pulse through a dual exploding wire array \cite{suttle2016structure,hare2018experimental,lebedev2019exploring}, and either observed plasmoid formation with minimal cooling at higher Lundquist numbers  $S_L \sim 100$ \cite{hare2017anomalous}, or evidence for sudden cooling of the ions at a lower $S_L <10$ \cite{suttle2018ion}. 

In this Letter, we present results from the Magnetic Reconnection on Z (MARZ) experiments, which generate a radiatively-cooled reconnection layer by driving a dual exploding wire array using the Z machine (20 MA, 300 ns rise time, Sandia National Labs) \cite{sinars2020review}. In contrast to previous pulsed-power experiments, the MARZ experiments demonstrate both a high $S_L \sim 100$ and a high cooling parameter ($R_{\text{cool}} > 100$). We make the first quantitative measurements of reconnection in a strongly radiatively-cooled regime, using temporally- and spatially-resolved X-ray diagnostics to measure emission from the reconnection layer. This is of particular astrophysical significance, as radiative emission is the key, and often only, signature of reconnection in extreme astrophysical objects \cite{uzdensky2011magneticb}. We observe the formation and subsequent radiative cooling of the reconnection layer. Furthermore, the layer exhibits sub-millimeter-scale fast-moving hotspots that emit most of the high-energy X-rays from the layer. Radiative resistive magnetohydrodynamic (MHD)  simulations of the experiment \cite{datta2024simulations} show that these hotspots are likely to be plasmoids generated by the tearing instability \cite{uzdensky2010fast}. These simulations are performed in GORGON --- an Eulerian resistive MHD code with van Leer advection \cite{ciardi2007evolution} --- which implements $P_{1/3}$ multi-group radiation transport, using spectral emissivity and opacity data from SpK \cite{crilly2022spk}. Simulation details are provided in Ref. \cite{datta2024simulations}.

\begin{figure}
\includegraphics[page=1,width=0.48\textwidth]{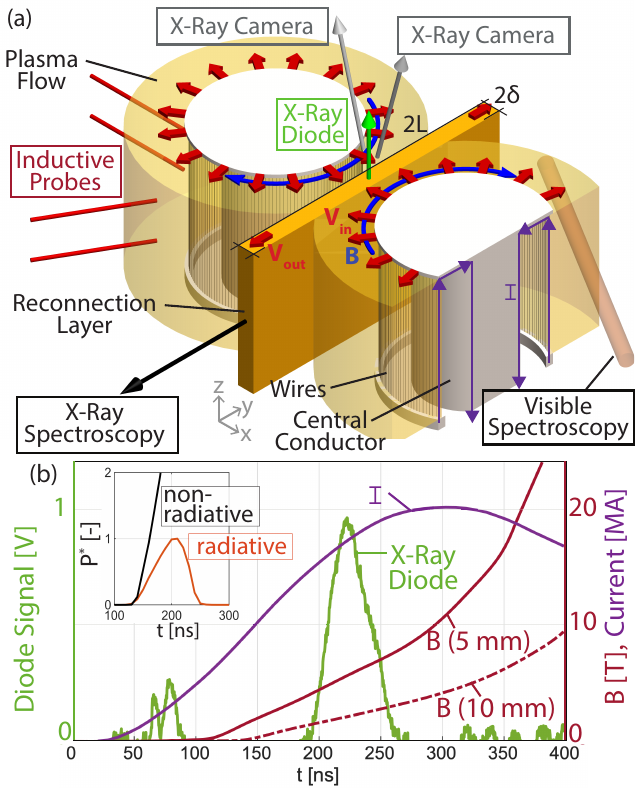}
\centering
\caption{
(a) A three-dimensional model of the dual wire array load hardware and diagnostic setup. Each array generates radially diverging plasma flows with azimuthal magnetic field lines, that collide to generate a reconnection layer in the midplane. (b) Temporal variation of the current (purple),  X-ray diode signal (green), and magnetic field at 5~mm and 10~mm from the wires (red). Inset: Simulated X-Ray emission from the reconnection layer, filtered with \SI{8}{\micro\meter} Be, for the non-radiative (black) and the radiatively-cooled (orange) cases. $P^*$ is the X-ray power normalized using the peak power in the radiative simulation.}
\label{fig:load}
\end{figure}

Fig.~\ref{fig:load}a shows the experimental setup. The load comprises two $\SI{40}{\milli \meter}$ tall, $\SI{40}{\milli \meter}$ diameter exploding wire arrays, each with 150 equally-spaced $\SI{75}{\micro \meter}$ diameter aluminum wires. The arrays have a $\SI{60}{\milli \meter}$ center-to-center separation, with a $\SI{10}{\milli \meter}$ gap between the mid-plane and the wires. Each array generates radially diverging plasma flows with azimuthally-oriented frozen-in magnetic field that is advected with the flow \cite{suttle2016structure,suttle2018ion,hare2017anomalous,hare2017formation,hare2018experimental}. The outflows collide at the mid-plane, where the anti-parallel magnetic field lines generate a reconnection layer. The arrays are driven in parallel by the Z machine; Fig.~\ref{fig:load}b shows the current $I(t)$ measured by a b-dot probe near the load \cite{webb2023radiation}, which is well approximated by $I(t) = \SI{20}{\mega\ampere} \sin^2\left(\pi/2 \times t/\SI{300}{\ns}\right)$. Photon Doppler velocimetry \cite{porwitzky2018uncertainties} measured equal division of current between the arrays across multiple shots. The arrays are over-massed, so they generate continuous plasma flows throughout the experiment without exploding \cite{harvey2009quantitative,datta2023plasma}. The plasma flows are highly-collisional ($\lambda_{ii} \approx 1-\qty{10}{\nano \meter}$), strongly-driven ($M_A \approx 7$, $M_S > 10$), and quasi-2D with minimal variation in the $z$ direction. There is no guide field, and the ion skin depth ($d_i \approx \SI{0.1}{\milli \meter}$) is small compared to the width of the reconnection layer ($\sim \SI{1}{\milli \meter}$). 

We use inductive probes and visible spectroscopy to characterize the magnetic field, ion density, and electron temperature in the outflows from the arrays, which form the inflows into the reconnection layer. The inductive probes \cite{byvank2017applied} are positioned at different radii (\SI{5}{\milli \meter} and \SI{10}{\milli \meter}) around the arrays (see Fig.~\ref{fig:load}a). Opposite-polarity probe pairs with a $\SI{1}{\centi \meter}$ vertical separation are used at each location for redundancy and to check for the influence of the common mode \cite{datta2022time,datta2022structure}. The probes are calibrated, and the magnetic field is determined by numerically integrating the signals. The magnetic fields $B(t)$ measured by probes at $\SI{5}{\milli \meter}$ and $\SI{10}{\milli \meter}$ from the wires (see Fig.~\ref{fig:load}b) are similar in shape, but displaced in time, consistent with the advection of the field between the probes at a flow velocity of $140 \pm \SI{30}{\kilo \meter \per \second}$ \cite{datta2022time}.

We use an optical fiber to collect visible radiation from the plasma along a path (diameter $\approx \SI{4}{\milli \meter}$) in the $xy$ plane, centered $\SI{8}{\milli \meter}$ from the wires, to a spectrometer ($400-\SI{700}{\nano \meter}$ range, $\SI{0.3}{\nano \meter}$ resolution) coupled to a streak camera ($\SI{600}{\nano \second}$ sweep time, $\SI{0.3}{\nano \second}$ resolution) \cite{schaeuble2021experimental}. The spectra show Al-II (Mg-like) and Al-III  (Na-like) emission lines. We infer density from the width of the well-isolated Al-II \SI{466.4}{\nano \meter} line, and temperature from the line ratio of the inter-stage Al-II \SI{466.4}{\nano \meter} and Al-III (\SI{448.1}{\nano \meter}, \SI{452.4}{\nano \meter}) lines. This is done by fitting synthetic spectra, generated using PrismSPECT and radiation transport simulations, to the experimental data \cite{datta2023machine}. The electron temperature in the outflows from the arrays increases from $\SI{1.8\pm0.4}{\electronvolt}$ to $\SI{2.1\pm0.4}{\electronvolt}$ between $200-\SI{240}{\nano \second}$, and the ion density is $5-\SI{8e17}{\per \centi \meter \cubed}$.

A filtered X-ray diode viewing the reconnection layer from the top (vertical green arrow in Fig.~\ref{fig:load}a) provides time-resolved measurements of the emitted X-ray power. The $\SI{8}{\micro \meter}$ beryllium filter transmits photons with energy $\SI{>1}{\kilo\electronvolt}$. X-ray emission from the reconnection layer (green curve in  Fig.~\ref{fig:load}b) exhibits a sharp peak at $\SI{220}{\nano\second}$, with a full-width-at-half-maximum of about $\SI{50}{\nano \second}$. The signal is reproducible over multiple shots and viewing angles. The X-ray emission is narrower than the driving current pulse, and reaches a maximum well before peak current, indicating that this feature is driven by the dynamics of the reconnection layer, rather than the driving current.

\begin{figure}[t!]
\includegraphics[page=2,width=0.5\textwidth]{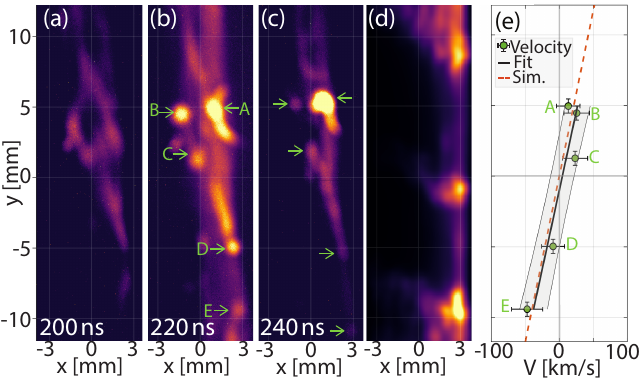}
\centering
\caption{
(a-c) Time-gated X-ray images ($\SI{10}{\nano \second}$ exposure) of the reconnection layer at $\SI{200}{\nano \second}$, $\SI{220}{\nano \second}$, $\&$ $\SI{240}{\nano \second}$, showing hotspots (green arrows) inside a bright elongated layer. (d) A synthetic X-ray image of the layer obtained by post-processing a 3D simulation at $\SI{220}{\nano \second}$ with X-ray emission and radiation transport modeling \cite{datta2024simulations}. (e) Velocity of the hotspots inside the reconnection layer. The solid line is a linear fit to the hotspot velocity, and dashed orange line is the simulated outflow velocity in the reconnection layer, averaged between $200-\SI{250}{\nano \second}$.
}
\label{fig:UXI}
\end{figure}

We probe the temporal evolution of the reconnection layer with two ultra-fast X-ray imaging pinhole cameras \cite{claus2015overview}. The cameras (4 frames, $\SI{10}{\nano \second}$ exposure) provide a $25 \times \SI{12.5}{\milli \meter \squared}$ field of view through a $\SI{500}{\micro \meter}$ diameter pinhole (magnification = $1\times$, geometric resolution $\approx \SI{1}{\milli \meter}$), filtered with $\SI{2}{\micro \meter}$-thick aluminized mylar ($\SI{>100}{\electronvolt}$ photons). The cameras view the reconnection layer with polar angles of $\theta= \SI{9}{\degree}$ and $\theta = \SI{12}{\degree}$ respectively, and with azimuthal angles (from the $x-$axis) of $ \phi = \SI{170}{\degree}$ and $\phi = \SI{40}{\degree}$, thus viewing both the top and side of the layer (see Fig.~\ref{fig:load}a). Figs.\ \ref{fig:UXI} (a-c) show an elongated, bright layer with strongly-emitting, localized ($\sim \SI{1}{\milli \meter}$ size) hotspots (indicated by green arrows) between $200 - \SI{240}{\nano \second}$. The intensity of the emission initially increases, reaching a maximum at $\SI{220}{\nano \second}$, and then decreases. Fig.~\ref{fig:UXI} shows images from only one camera ($\theta = \SI{12}{\degree},  \,  \phi = \SI{40}{\degree}$); images from the second camera, containing similar features, are in the supplementary material. The hotspots move along the $y$-direction, away from the center of the layer. From the translation of the hotspots between $220 - \SI{240}{\nano \second}$, we estimate their velocity (Fig.~\ref{fig:UXI}e), which shows acceleration from rest to $50\pm\SI{20}{\kilo \meter \per \second}$ over $\SI{10}{\milli \meter}$.  We show later that this is consistent with the expected and simulated outflow velocity from the reconnection layer. 

\begin{figure}[b!]
\includegraphics[page=4,width=0.5\textwidth]{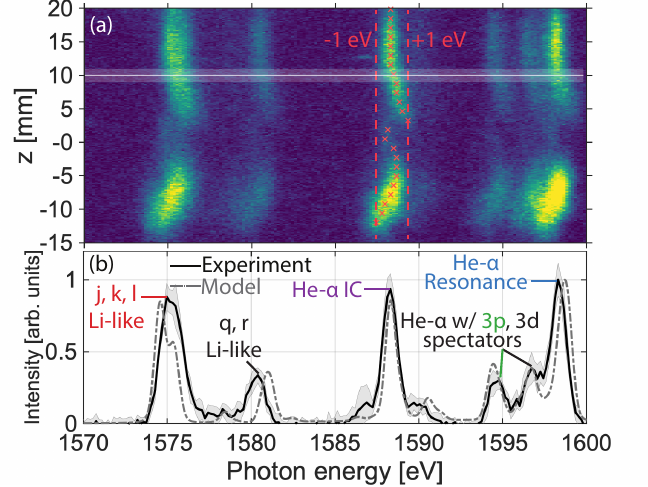}
\centering
\caption{(a) Time-integrated X-ray spectra from the reconnection layer showing Al K-shell emission. Red crosses show the axial variation in the position of the He-$\alpha$ inter-combination line. (b) X-ray spectrum averaged over $z = \SI{10 \pm 1}{\milli \meter}$ [white line in (a)] showing He-like and Li-like satellite transitions (black), and calculated spectrum from the SCRAM and radiation transport model for 170 eV, \SI{1e18}{\per \centi \meter \cubed} hotspot density, and \SI{0.5}{\milli \meter} hotspot size (grey dashed).}
\label{fig:XRS3}
\end{figure}

An X-ray spectrometer with a spherically-bent crystal \cite{harding2015analysis} provides time-integrated spatially-resolved (along $z$, resolution: $\Delta z \approx \SI{200}{\micro \meter}$) spectral measurements (resolution: $\Delta E \approx 0.5$ eV) of X-ray emission from the reconnection layer (see Fig.~\ref{fig:load}a). Fig.~\ref{fig:XRS3}a shows the X-ray spectrum, which exhibits He-like and Li-like satellite Al K-shell transitions with energies of about 1.6 keV. These transitions are labeled in Fig.~\ref{fig:XRS3}b, which shows a lineout of the recorded spectrum averaged over $z = 10\pm\SI{1}{\milli \meter}$. Here, we show the X-ray spectrum from the same experimental shot as the diode signal and X-ray images; however, this spectrum is reproducible across multiple shots. Although the spectrum is time-integrated, the filtered X-ray diode ($\SI{8}{\micro \meter}$ Be, $>1$ keV) localizes the spectra in time to $220 \pm \SI{25}{\nano \second}$. 

\begin{table*}\centering
\ra{1.3}
\caption{Plasma parameters in the inflow region and reconnection layer at time of onset of radiative cooling (220 ns). Values in bold are measured experimentally, while other values are estimated/inferred values. Parenthetical values show bounds from X-ray spectroscopy.}
\begin{tabular}{ccccccccccccc}
\hline
 & $n_i \, [\times \SI{e18}{\per \centi \meter \cubed}$] & $T_e$ [eV] & $\bar{Z}$ & $B_y$ [T]  & p [MPa] & $V_x$ & $V_y$ & $V_A$ & $C_S$ [km/s] & $d_i$ [mm] & $S_L$ & $\tau_{E}$ [ns] \\
\hline
Pre-shock Inflow & {\bf 0.8} & {\bf 1.9} & {\bf 2} & {\bf 3.9} & 0.7 & {\bf 140} &  - &  20 & 5 & 0.7 & -  & 2\\
Post-shock Inflow & 6& 30 & 8 & 30 & 300 & 20 & - &  50& 30 & 0.1 & - & 1\\
Reconnection Layer & 6 $(\mathbf{\lesssim  5})$ & 60 $(\mathbf{\lesssim 75})$ & 10 & - & 700 & - & {\bf 72} & -& 50 & 0.1  & 120 & 1 \\
\hline
\end{tabular}
\label{tab:table}
\end{table*}


Combining the spatial, temporal, and spectral measurements of the X-ray emission provides information about the evolution of the reconnection layer. The narrow burst of X-ray emission from the reconnection layer, as recorded simultaneously by the diode (Fig.~\ref{fig:load}b) and X-ray cameras (Fig.~\ref{fig:UXI}), provides evidence for formation and radiative cooling of the layer. The initial rise of the diode signal indicates increasing density and/or temperature of the layer during the formation stage. The layer temperature is initially hot enough to generate $\SI{>1}{\kilo\electronvolt}$ emission. The sharp fall in the X-ray emission after \SI{220}{\nano \second} is consistent with a rapidly cooling reconnection layer. X-ray images of the layer exhibit a similar intensity evolution as the diode, with a maximum intensity around $\SI{220}{\nano\second}$ (Fig.~\ref{fig:UXI}). Resistive MHD simulations confirm that X-ray emission from the reconnection layer provides a diagnostic signature of strong radiative cooling \cite{datta2024simulations}: in the absence of radiative cooling, simulations show that X-ray emission $(> \SI{1}{\kilo \electronvolt})$ would continue to rise due to increasing layer density at a constant, high temperature, as shown by the black curve in the inset of Fig.~\ref{fig:load}b. However, radiative cooling results in a sharp decline in the simulated X-ray emission after an initial rise (orange curve, Fig.~\ref{fig:load}b), similar to the experimentally-measured emission.

To obtain quantitative measurements of temperature and density in the reconnection layer from the X-ray spectra in Fig.~\ref{fig:XRS3} during the onset of radiative cooling, we use non-local thermodynamic equilibrium spectral emissivity and absorption opacity tables generated using SCRAM \cite{hansen2007hybrid} to solve the radiation transport equation \cite{Drake2006} along the diagnostic line-of-sight. SCRAM includes spectral line broadening effects, and incorporates photo-pumping by assuming cylindrical geometry with diameter of $\SI{1}{\milli \meter}$. These radiation transport calculations constrain the composition of the emitting plasma: because of the higher opacity of the He-$\alpha$ resonance transition compared to the other lines in Fig.~\ref{fig:XRS3}b, it is strongly damped in homogeneous layers of length $>\SI{1}{\milli \meter}$. Therefore, only sub-millimeter-sized hotspots, strongly emitting in the Al K-shell, can account for the similar measured intensities of the He-$\alpha$ resonance and inter-combination (IC) lines. The experimental spectrum is best matched by the emission from localized, dense hotspots of size $<\SI{1}{\milli \meter}$ and $T_e \approx \SI{170 \pm 30}{\electronvolt}$, embedded within a colder, less dense reconnection layer. An example spectrum calculated assuming emission from a single hotspot of size $d = \SI{0.5}{\milli \meter}$, temperature $T_e = \SI{170}{\electronvolt}$, and ion density $n_i = \SI{1e18}{\per \centi \meter \cubed}$ in a non-emitting ($\epsilon = 0$) non-absorbing layer ($\alpha = 0$), is shown by the dashed curve in Fig.~\ref{fig:XRS3}b, and reproduces the line ratios and line widths of the experimental spectrum. The model includes the effect of source and instrument broadening, but neglects Doppler shift, which is $\SI{<0.25}{\electronvolt}$, as calculated from the hotspot velocities in Fig.~\ref{fig:UXI}e. Radiation transport also provides an upper limit of about $\SI{75}{\electronvolt}$ on the layer temperature, to prevent over-damping of the He-$\alpha$ resonance line generated from the hotspots. This model is further supported by X-ray images of the layer (see Fig.~\ref{fig:UXI}), which show brightly-emitting hotspots $<\SI{1}{\milli \meter}$ in size inside the less brightly-emitting reconnection layer. Fig.~\ref{fig:XRS3}a further shows that the hotspots form contiguous elongated structures along $z$ of length $\sim \SI{10}{\milli\meter}$.

To constrain the density ($n_i$), temperature ($T_e$), and size ($d$) of the hotspots for which the experimental X-ray spectrum is valid, we uniformly and randomly sample values of $n_i, \, T_e, \text{ and } d$ to obtain solutions that match within $20\%$ the experimentally observed line ratios. Comparing the relative intensities of the He-$\alpha$ IC with the Li-j satellite line and  He-$\alpha$ with 3p spectator transition constrains $T_e$ to a narrow band around $\SI{170\pm 30}{\electronvolt}$, and provides an upper bound of $\sim \SI{5e18}{\per \centi \meter \cubed}$ on the hotspot ion density. Assuming that the hotspot density lies between the upper bound of $n_i \leq \SI{5e18}{\per \centi \meter \cubed}$ and the lower bound of $n_i \geq \SI{5e17}{\per \centi \meter \cubed}$ (inflow density from visible spectroscopy), this sampling contrains $0.3 \leq d \leq \SI{0.5}{\milli \meter}$.

We also observe hotspots in synthetic X-ray images produced from 3D resistive MHD simulations of the experiment \cite{datta2024simulations} (Fig.~\ref{fig:UXI}d). In these simulations, plasmoids generated by the tearing instability are hotter and denser than the rest of the current sheet, and are regions of intense Al K-shell X-ray emission \cite{datta2024simulations}. The experimentally measured hotspot velocity is also consistent with the simulated velocity in the outflows from the reconnection layer, as seen in Fig.~\ref{fig:UXI}e. The plasmoid instability in the simulations occurs at Lundquist numbers smaller than the canonical critical value $S_L^* \sim \SI{e4}{}$ \cite{uzdensky2010fast}; which may be due to density perturbations, compressibility, or non-uniform plasma resistivity \cite{datta2024simulations}. The simulations further show that despite the generation of plasmoids, the reconnection rate follows the scaling predicted by the radiatively-cooled Sweet-Parker theory in Ref.\ \cite{uzdensky2011magnetic}.

These 3D simulations also show that the plasmoid position in the $xz$-plane varies along the $z$ direction, due to the MHD kink instability \cite{datta2024simulations}. We see preliminary evidence for this modulation in the axially-resolved X-ray spectrum in Fig.~\ref{fig:XRS3}a, where the spectral positions of the lines exhibit modulations of up to $\SI{1}{\electronvolt}$ along $z$. Ray tracing calculations \cite{harding2015analysis} show that these deviations correspond to $\SI{1}{\milli \meter}$ $x$-displacements in the source position, comparable to the kink instability amplitude in the simulation.

Using these experimental measurements, we calculate key physical and dimensionless parameters for the inflows and the reconnection layer. Table \ref{tab:table} summarizes the experimentally determined parameters in the layer inflows (array outflows) right before onset of radiative cooling ($\SI{220}{\nano\second}$). The magnetic field is averaged from probe measurements at $5$ and $\SI{10}{\milli \meter}$, while $n_i$ and $T_e$ are from visible spectroscopy. From these, we estimate the thermal pressure $p$, adiabatic index $\gamma \approx 1.2$ \cite{Drake2006}, and the Alfvén $V_A = B/\sqrt{\mu_0\rho}$ and sound speeds $C_S = \sqrt{\gamma p/\rho}$. We assume $T_i \approx T_e$ as the estimated energy equilibration time $\tau_E \approx \SI{2}{\nano \second}$ \cite{richardsonnrl2019} is smaller than the hydrodynamic time $\tau_{\text{hydro}} = L/V  \approx \SI{70}{\nano \second}$. 

The layer inflows are super-Alfvénic ($M_A \approx 7$); consequently we expect magnetic flux pile-up to generate shocks upstream of the reconnection layer, dividing the inflow to the reconnection layer into pre-shock and post-shock regions, as observed in both the simulations \cite{datta2024simulations} and previous experiments \cite{suttle2016structure,suttle2018ion,fox2011fast,olson2021regulation}. Since the plasma is highly collisional, we estimate the post-shock plasma conditions by solving the Rankine-Hugoniot equations for a fast perpendicular MHD shock \cite{goedbloed_keppens_poedts_2010,boyd_sanderson_2003}, resulting in an increase in the density and magnetic field, and a decrease in the inflow velocity by a factor of about 8. Estimated values in the post-shock inflow region are shown in Table \ref{tab:table}. 
 
To estimate the plasma parameters in the reconnection layer, we assume that (1) a pressure balance exists between the layer and the post-shock inflow, and (2) right before onset of cooling, there is little compression of the layer, such that mass density is roughly equal inside and just outside the layer in the post-shock inflow. Both assumptions are supported by simulations \cite{datta2024simulations}. The estimated layer temperature and ion density at this time are therefore $\SI{60}{\electronvolt}$ and $\SI{6e18}{\per \centi \meter \cubed}$ respectively. The temperature is below the upper bound ($T_e \lesssim \SI{75}{\electronvolt}$), while the density is close to the upper bound ($n_i \lesssim  \SI{5e18}{\per \centi \meter \cubed}$) determined from X-ray spectroscopy. 

We extrapolate the linear velocity trend in Fig.~\ref{fig:UXI}e to $y = L$ ($L = \SI{15}{\milli \meter}$, $0.5\times$ field line radius of curvature at the mid-plane), and estimate the layer outflow velocity $V_{out} \approx \SI{72}{\kilo \meter \per \second}$. 
This velocity closely matches the magnetosonic velocity $V_{MS} = (V_{A,in}^2 + C_{S,L}^2)^{0.5} \approx \SI{70}{\kilo \meter \per \second}$ (computed from the Alfvén speed outside the layer $V_{A,in}$, and the sound speed inside the layer $C_{S,L}$), which is the theoretical outflow velocity from the reconnection layer \cite{ji1999magnetic,hare2017anomalous}. The estimated Lundquist number is $S_L = L V_A/\bar{\eta}\approx 120$, and 
the predicted Sweet-Parker layer width is $\delta_{SP} \approx L (S_L)^{-1/2} \approx \SI{1.4}{\milli \meter}$ \cite{parker1957sweet,yamada2010magnetic}. This width is much larger than both the estimated ion-ion mean free path ($\lambda_{ii} \approx \SI{2}{\nano \meter}$) and the ion skin depth ($d_i \approx \SI{0.1}{\milli \meter}$), indicating high collisionality, and justifying the use of resistive MHD models. Using the post-shock inflow velocity $V_{in} \approx \SI{20}{\kilo \meter \per \second}$, we infer the reconnection rate at this time $V_{in}/V_{out} \approx 0.3$, which is roughly comparable to the Sweet-Parker rate ${S_L^{-1/2}\approx0.1}$ \cite{parker1957sweet}.

Finally, we provide order-of-magnitude estimates of the dominant terms in the layer power balance. The inferred current density $j \sim B_{in} / (\delta_{SP} \mu_0) \approx \SI{20}{\giga \ampere \per \meter \squared}$ in the layer, provides an Ohmic heating power of $P_{\Omega} \sim \SI{e15}{\watt \per \meter \cubed}$. The estimated compressional heating is $P_{\text{comp}} \sim pV_{in}/\delta_{SP} \sim \SI{e16}{\watt \per \meter \cubed}$. Viscous heating and thermal conduction are negligible. 
We estimate the radiative loss from the layer by solving the radiation transport equation along a mean chord with a length approximated as the volume-to-surface area ratio of a rectangular slab of width $2\delta$ and length $2L$ \cite{datta2024simulations}. Using spectral emissivities and opacities from SpK \cite{crilly2022spk}, the resulting loss rate is $P_{\text{rad}} \sim \SI{e18}{\watt \per \meter \cubed}$, corresponding to a cooling parameter $R_{\text{cool}} \sim 400$ at 220 ns. The experimentally-measured hotspot temperature and density provide an upper bound on the cooling rate $P_\text{{rad, hotspot}} \sim \SI{e19}{\watt \per \meter \cubed}$. Cooling therefore dominates heating within the reconnection layer, consistent with the strong cooling observed in Figs.\ \ref{fig:load}(b) \& \ref{fig:UXI}(a-c).

In summary, we present the first experimental evidence of strong radiative cooling in a pulsed-power-driven reconnection experiment with $S_L>100$. The key results are:\\
1. The reconnection layer exhibits millimeter-scale fast-moving hotspots with strong X-ray emission, consistent with the presence of magnetic islands generated by the plasmoid instability in 3D resistive MHD simulations. \\
2. The majority of the high-energy X-rays are generated by these hotspots, which exhibit a temperature (about \SI{170}{\electronvolt}) higher than both the inflow (about \SI{2}{\electronvolt}) and bulk layer temperature (\SI{<75}{\electronvolt}). \\
3. The reconnection layer undergoes strong radiative cooling, characterized by the rapid decrease in X-ray emission from the layer. 

Using experimentally measured values in the inflow region, we estimate the plasma parameters inside the reconnection layer, which are consistent with the bounds determined from experimental diagnostics. Lastly, we estimate that at the time of peak X-ray emission, the cooling rate in the layer is much higher than the heating rate. Strong cooling is necessary to trigger radiative collapse of the reconnection layer. Future experiments will characterize the evolution of the plasma properties during this radiative collapse process, using time-resolved measurements of the layer width, temperature, density, and outflow velocity. The findings in this Letter are of particular relevance to the generation of radiative emission from reconnection-driven astrophysical events, and to the global dynamics of reconnection in strongly-cooled systems. These experiments also provide a novel platform for the investigation of radiative effects in HED and laboratory astrophysics experiments, and for validation of radiation (magneto) hydrodynamic and atomic codes.
\newline
\newline
The authors would like to thank the Z machine operations teams and the target fabrication team for their contributions to this work.
Experimental time on the Z facility was provided through the Z Fundamental Science Program.
This work was funded by NSF and NNSA under grant no. PHY2108050, and by the NSF EAGER grant no. PHY2213898. RD acknowledges support from the MIT MathWorks and the MIT College of Engineering Exponent fellowships. DAU gratefully acknowledges support from NASA grants 80NSSC20K0545 and 80NSSC22K0828. This work was supported by Sandia National Laboratories, a multimission laboratory managed and operated by National Technology and Engineering Solutions of Sandia, LLC, a wholly owned subsidiary of Honeywell International Inc., for the U.S. Department of Energy’s National Nuclear Security Administration under contract DE-NA0003525. This paper describes objective technical results and analysis. Any subjective views or opinions that might be expressed in the paper do not necessarily represent the views of the U.S. Department of Energy or the United States Government.

\setcitestyle{square}
\bibliography{main}
\setcitestyle{square}

\pagebreak

\appendix

\begin{figure*}[h!]
\includegraphics[width=0.8\textwidth]{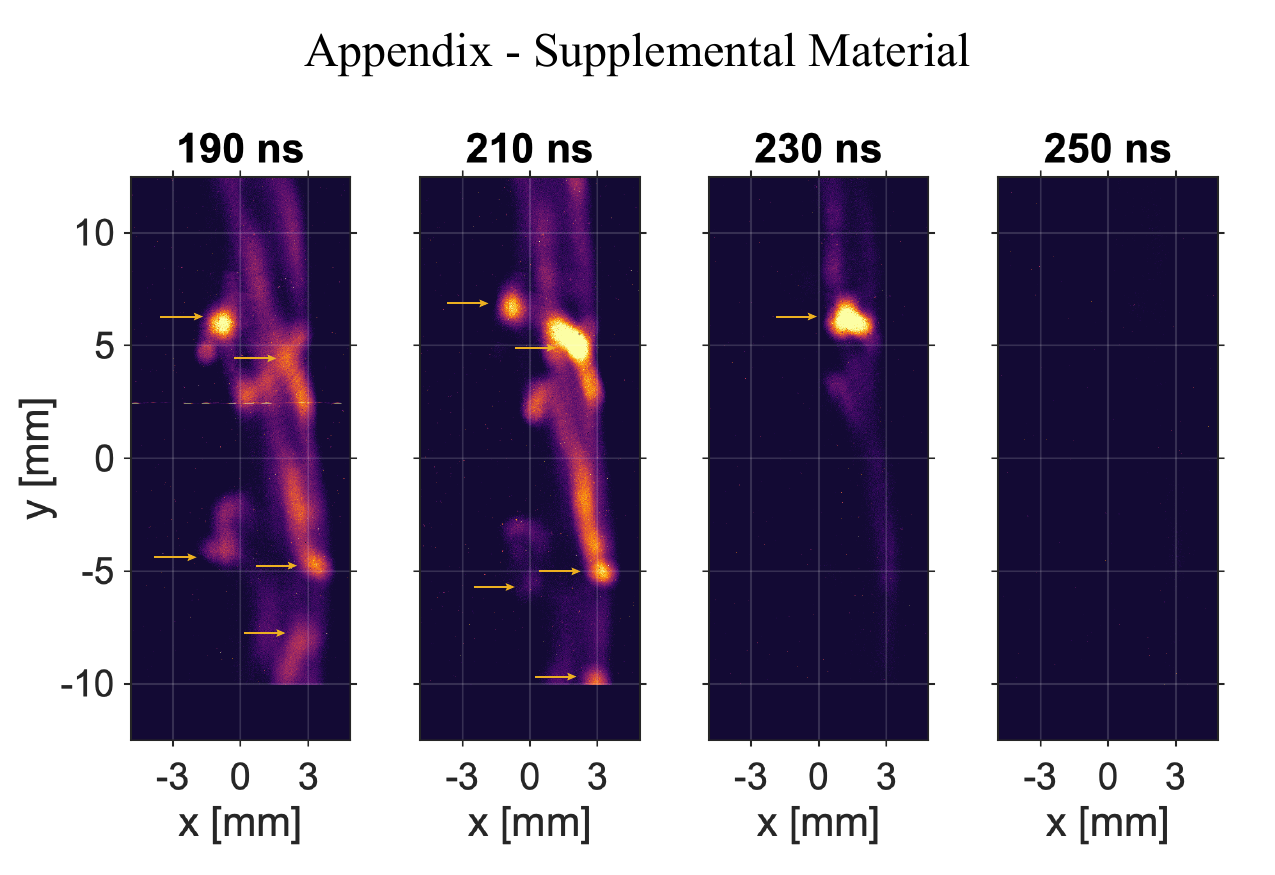}
\centering
\caption{Time-gated X-ray images (10 ns exposure) of the reconnection layer at 190-250 ns, showing hotspots (yellow arrows) inside a bright elongated layer. The pinhole camera is filtered with 2um-thick aluminized Mylar, and views the reconnection layer with polar angle $\theta =\SI{9}{\degree}$ and azimuthal angle $\phi =\SI{170}{\degree}$.}
\end{figure*}

\end{document}